\documentclass[two column,linenumbers,superscriptaddress,groupedaddress amssymb, amsmath,nobibnotes,nofootinbib, aps, showpacs, prb]{revtex4}
\usepackage[dvips]{graphicx}
\usepackage{graphicx}
\usepackage{lineno}

\begin{document}
%\linenumbers
\title{Searching for triplet superconductivity in the Quasi-One-Dimensional  K$_2$Cr$_3$As$_3$}
\author{D. T. Adroja}
\email{devashibhai.adroja@stfc.ac.uk}
\affiliation{ISIS Facility, Rutherford Appleton Laboratory, Chilton, Didcot Oxon, OX11 0QX, United Kingdom} 
\affiliation{Highly Correlated Matter Research Group, Physics Department, University of Johannesburg, PO Box 524, Auckland Park 2006, South Africa}
\author{A. Bhattacharyya}
\email{amitava.bhattacharyya@stfc.ac.uk}
\affiliation{ISIS Facility, Rutherford Appleton Laboratory, Chilton, Didcot Oxon, OX11 0QX, United Kingdom} 
\affiliation{Highly Correlated Matter Research Group, Physics Department, University of Johannesburg, PO Box 524, Auckland Park 2006, South Africa}
\author{M. Telling}
\affiliation{ISIS Facility, Rutherford Appleton Laboratory, Chilton, Didcot Oxon, OX11 0QX, United Kingdom} 
\author{Yu. Feng}
\affiliation{State Key Laboratory of Surface Physics and Department of Physics, Fudan University, Shanghai 200433, China}
\author{M. Smidman}
\affiliation{Center for Correlated Matter and Department of Physics, Zhejiang University, Hangzhou 310058, China}
\affiliation{Collaborative Innovation Center of Advanced Microstructures, Nanjing University, Nanjing 210093, China}
\author{B. Pan}
\affiliation{State Key Laboratory of Surface Physics and Department of Physics, Fudan University, Shanghai 200433, China}
\author{J. Zhao}
\affiliation{State Key Laboratory of Surface Physics and Department of Physics, Fudan University, Shanghai 200433, China}
\author{A. D. Hillier}
\affiliation{ISIS Facility, Rutherford Appleton Laboratory, Chilton, Didcot Oxon, OX11 0QX, United Kingdom} 
\author{F.  L. Pratt}
\affiliation{ISIS Facility, Rutherford Appleton Laboratory, Chilton, Didcot Oxon, OX11 0QX, United Kingdom} 
\author{A. M. Strydom}
\affiliation{Highly Correlated Matter Research Group, Physics Department, University of Johannesburg, PO Box 524, Auckland Park 2006, South Africa}

\date{\today}

\begin{abstract}
%\linenumbers
The superconducting state of the newly discovered superconductor K$_2$Cr$_3$As$_3$ with a quasi-one-dimensional crystal structure ($T_{\bf c}\sim$ 6 K) has been investigated  by using magnetization and muon-spin relaxation or rotation ($\mu$SR) measurements. Our analysis of the temperature dependence of  the superfluid density obtained from  the transverse field (TF) $\mu$SR  measurements fit very well to an isotropic $s$-wave character for the superconducting gap. Furthermore a similarly good fit can also be obtained using a $d$-wave model with line nodes. Our zero-field $\mu$SR measurements do reveal very weak evidence of the spontaneous appearance of an internal magnetic field near the transition temperature, which might indicate that the superconducting state is not conventional. This observation suggests that the  electrons are paired via unconventional channels such as spin fluctuations, as proposed on the basis of theoretical models of K$_2$Cr$_3$As$_3$. Furthermore,  from our TF $\mu$SR study the magnetic penetration depth $\lambda_L$, superconducting carrier density $n_s$, and effective-mass enhancement $m^*$ have been estimated to be $\lambda_L(0)$ = 454(4) nm, $n_s$ = 2.4$\times$10$^{27}$ carriers/m$^3$, and $m^*$ = 1.75 $m_e$, respectively.
\end{abstract}


\maketitle

The superconducting gap structure of strongly correlated f$-$ and d$-$ electron superconductors is very important in understanding
the physics of unconventional pairing mechanism in this class of materials. The recently discovered superconductors with a quasi-one-dimensional (Q1D) crystal structure, K$_2$Cr$_3$As$_3$ T$_{\bf c}\sim$ 6.1 K, Rb$_2$Cr$_3$As$_3$ T$_{\bf c}\sim$ 4.8 K and Cs$_2$Cr$_3$As$_3$ T$_{\bf c}\sim$ 2.2 K have been intensively investigated both experimentally and theoretically~\cite{J. Bao,T. Kong,Z. Tang,Z. Tang1,G. M. Pang,X. Wu,H. Z.} as they are strong candidates for a multiband triplet pairing state. Searching for triplet superconductivity (SC) has been one of major research efforts recently partly due to its intrinsic connection to topologically related physics and quantum computations. These new superconductors are conjectured to possess an unconventional pairing mechanism~\cite{J. Bao, Z. Tang1,G. M. Pang,X. Wu,H. Z.}. There are several experimental evidences. Firstly, its upper critical field $H_{c2}$ is 3 times higher than the Pauli limit, indicating that the BCS-type pairing is unfavorable ~\cite{T. Kong}. Secondly, strong electronic correlations which are a common feature of unconventional superconductivity were revealed by a large electronic specific heat coefficient and non-Fermi liquid transport behavior ~\cite{J. Bao}. This is consistent with the the Q1D crystalline structure of A$_2$Cr$_3$As$_3$ (A = K, Rb and Cs) and represent a possible realization of Luttinger liquid state ~\cite{H. Z. Zhi}. Thirdly, line nodal gap symmetry was revealed by London penetration depth measurements of K$_2$Cr$_3$As$_3$ ~\cite{G. M. Pang}. 

\begin{figure}[t]
\centering
\includegraphics[width = \linewidth]{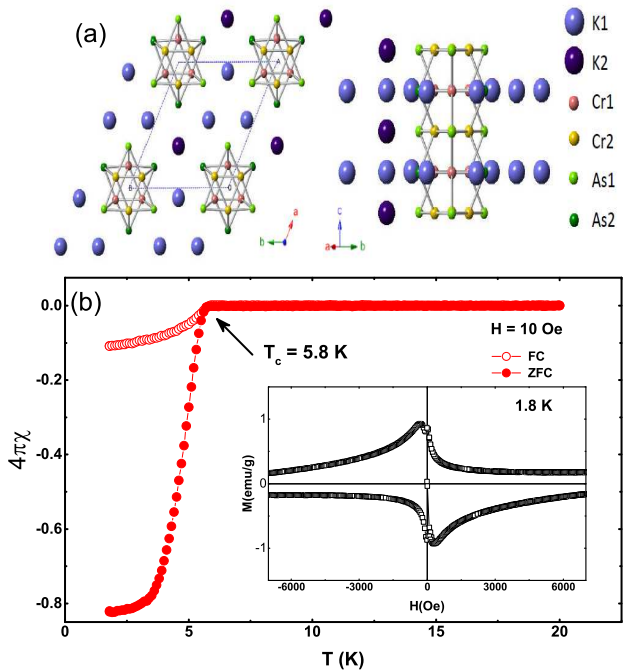}
\caption {(Color online) (a) The quasi-1D crystal structure of K$_2$Cr$_3$As$_3$ and (b) The temperature dependence of the dc magnetic susceptibility measured in zero-field cooled state (ZFC) and field cooled state (FC) of K$_2$Cr$_3$As$_3$ in the presence of an applied magnetic field of 10~G. The inset in (b) shows the magnetization vs field at 1.8 K.}
\end{figure}

Theoretically, by using DFT calculations, X. Wu {\it et. al.} predicted K$_2$Cr$_3$As$_3$ to be nearby a novel in-out co-plane magnetically ordered state and possess strong spin fluctuations~\cite{X. Wu,H. Z.}. Furthermore it has been shown that a minimum three-band model based on the $d_{z^2}$ , $d_{xy}$ and $d_{x^2-y^2}$ orbitals of one Cr sublattice can capture the band structures near the Fermi surfaces. In both weak and strong coupling limits, the standard random phase approximation (RPA) and mean-field solutions consistently yield the triplet $p_z-$wave pairing as the leading pairing symmetry for physically realistic parameters~\cite{X. Wu}. The triplet pairing is driven by the ferromagnetic fluctuations within the sublattice~\cite{X. Wu,H. Z.}. The gap function of the pairing state possesses line gap nodes on the $k_z$ = 0 plane of the Fermi surface. So it is highly likely that electrons are paired via unconventional channels such as spin fluctuations in K$_2$Cr$_3$As$_3$. NMR measurements indeed reveal the enhancement of spin fluctuations toward $T_{\bf c}$ in K$_2$Cr$_3$As$_3$~\cite{H. Z. Zhi}.  Furthermore  Y. Zhou {\it et. al.}~\cite{Y. Zhou} have shown theoretically that at small Hubbard U and moderate Hund's coupling, the pairing arises from the 3-dimensional (3D)  $\gamma$ band and has $f_{y(3x^2 -y^2)}$ symmetry, which gives line nodes in the gap function. At large U, a fully gapped $p$-wave
state dominates on the quasi-1D $\alpha$-band.

A polycrystalline sample of K$_2$Cr$_3$As$_3$ was prepared as discussed in Ref. [1]. A high quality powder sample of K$_2$Cr$_3$As$_3$ has been characterized using neutron diffraction and magnetic susceptibility. The magnetization data were measured using a Quantum Design Superconducting Quantum Interference Device magnetometer.  Muon spin relaxation ($\mu$SR) experiment were carried out on the MUSR spectrometer at the ISIS pulsed muon source of the Rutherford Appleton Laboratory, UK~\cite{sll}. The $\mu$SR experiments were conducted in zero$-$field (ZF), and transverse$-$field (TF) mode. A polycrystalline sample of K$_2$Cr$_3$As$_3$ was mounted in a sealed Titanium (99.99 \%) sample holder under the He-exchange gas, which was placed in a sorption cryostat that has a temperature range of 350 mK$-$50 K. Using an active compensation system the stray magnetic fields at the sample position were canceled to a level of 1 $\mu$T. TF$-\mu$SR experiments were performed in the superconducting mixed state in an applied field of 400 G, well above the lower critical field limit $H_{c1}$= 70 G of this material.~Data were collected in the  (a) field$-$cooled mode, where the magnetic field was applied above the superconducting transition and the sample was then cooled down to base temperature and (b) zero field cooled mode, where first the sample was cooled down to 2 K in ZF and then the magnetic field was applied. Muon spin relaxation is a dynamic method to resolve the type of the pairing symmetry in superconductors~\cite{js}. The mixed or vortex state in case of type-II superconductors gives rise to a spatial distribution of local magnetic fields; which demonstrates itself in the $\mu$SR signal through a relaxation of the muon polarization. The asymmetry of the muon decay in ZF is calculated by, $G_z(t) =[ {N_F(t) -\alpha N_B(t)}]/[{N_F(t)+\alpha N_B(t)}]$, where $N_B(t)$ and $N_F(t)$ are the number of counts at the detectors in the forward and backward positions and $\alpha$ is a constant determined from calibration measurements made in the paramagnetic state with a small (20 G) applied transverse magnetic field. The data were analyzed using the free software package WiMDA~\cite{FPW}.

\begin{figure}[t]
\vskip -0.0 cm
\centering
\includegraphics[width = 6 cm]{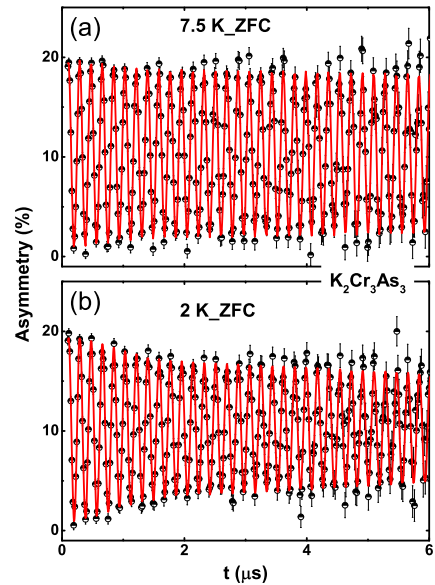}
\caption {(Color online) The transverse-field muon time spectra (one component) for K$_2$Cr$_3$As$_3$ collected (a) at $T$ = 7.5 K and (b) at $T$ = 2 K in a magnetic field $H$ = 400 G for zero-field cool (ZFC) state.}
\end{figure} 

\par

The analysis of the neutron powder diffraction (NPD) at 300 K reveals that the sample is single phase  and crystallizes with space group $P\bar{6}m2$ (No. 187). The hexagonal crystal structure obtained from NPD is shown in Fig. 1(a). The Q1D feature of K$_2$Cr$_3$As$_3$ is manifested by the chains of [Cr$_6$(As$_6$)]$_\propto$ octahedra running along the $c$ direction. Magnetic susceptibility measurement shows superconductivity occurs at 5.8 K and the superconducting volume fraction is  close to 100 \% at 2 K  (Fig. 1(b)), indicating bulk nature of superconductivity in  K$_2$Cr$_3$As$_3$. The magnetization $M(H)$ curve at 1.8 K (Inset in Fig.1(b)) shows a typical behaviour for type-II superconductivity.  The temperature dependent resistivity and variation of T$_{\bf c}$ with increasing field shows that upper critical field exceeds the Pauli limit by 3 times~\cite{J. Bao}.  In zero field, the temperature-dependent resistivity of K$_2$Cr$_3$As$_3$ is metallic~\cite{J. Bao}. Deviation from a linear temperature dependence is evident below 100 K and a $T^3$ dependence is roughly followed from just above T$_{\bf c}$ ($\sim$10 K) to $\sim$ 40 K.  At the superconducting transition, the specific heat jump is roughly 2.2~$\gamma$T$_{\bf c}$, which is larger than the simple $s-$wave BCS prediction 1.43~\cite{J. Bao}, possibly indicating strong coupling~\cite{J. P. Carbotte}.

\begin{figure}[t]
\vskip -0cm
\centering
\includegraphics[width = 7 cm]{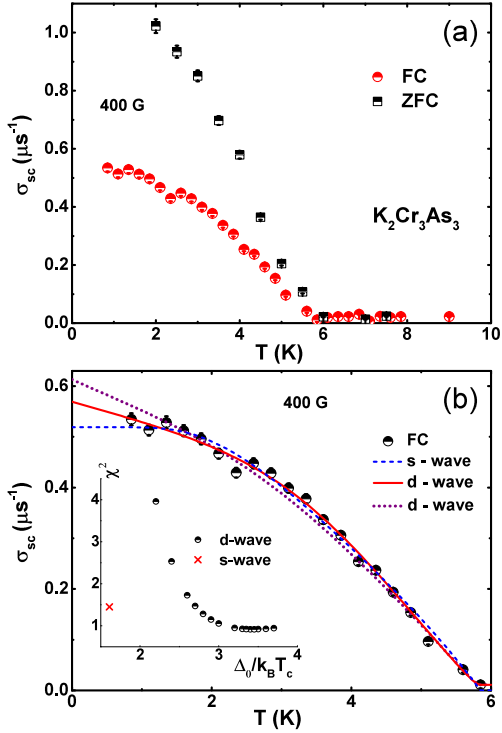}
\caption {(Color online) (a) The temperature dependence of  muon depolarization rate $\sigma_{sc}(T)$ of K$_2$Cr$_3$As$_3$ collected in an applied magnetic field of 400 G in zero-field cooled (ZFC) and field cooled (FC) modes. (b) $\sigma_{sc}(T)$ of FC mode (symbols) and the  lines are the fits to the data using  Eq. 2. The short-dash blue line shows the fit using an isotropic single-gap $s$-wave model with $\Delta_0$/$k_B$$T_{\bf c}$ = 1.6(1) and the solid red line and dotted purple line show the fit to $d$-wave model with  $\Delta_0$/$k_B$$T_{\bf c}$ = 3.2(2) and $\Delta_0$/$k_B$$T_{\bf c}$ = 2.40, respectively. The inset shows the plot of quality of fit $\chi^2$ vs $\Delta_0$/$k_B$$T_{\bf c}$}
\end{figure} 

\par

Fig. 2 (a) and (b) show the TF$-\mu$SR precession signals above and below $T_{\bf c}$ obtained in ZFC mode with an applied field of 400 G (well above $H_{c1}\sim$ 70 Oe but below $H_{c2}\sim$ 320 kOe). Below $T_{\bf c}$ the signal decays with time due to the inhomogeneous field distribution of the flux-line lattice. The TF$-\mu$SR asymmetry spectra were fitted using an oscillatory decaying Gaussian function,

\begin{equation}
\begin{split}
G_{z1}(t) = A_1\rm{cos}(2\pi \nu_1 t+\phi_1)\rm{exp}\left({\frac{-\sigma^2t^2}{2}}\right)\\ + A_2\rm{cos}(2\pi \nu_2 t+\phi_2),
\end{split}
\end{equation}

\noindent where $\nu_1$ and $\nu_2$ are the frequencies of the muon precession signal from the sample and from the background signal from the Ti-sample holder, respectively, $\phi_i$ ($i$ = 1,2) are the initial phase offsets.  The first term gives the total sample relaxation rate $\sigma$; there are contributions from both the vortex lattice ($\sigma_{sc}$) and nuclear dipole moments   ($\sigma_{nm}$), which is assumed to be constant over the entire temperature range below $T_{\bf c}$ [ where $\sigma$ = $\sqrt{(\sigma_{sc}^2+\sigma_{nm}^2)}$]. The contribution from the vortex lattice, $\sigma_{sc}$, was determined by quadratically subtracting the background nuclear dipolar relaxation rate obtained from spectra measured above $T_{\bf c}$. As $\sigma_{sc}$ is directly related to the magnetic penetration depth, it can be modeled by \cite{Prozorov}

\begin{equation}
\frac{\sigma_{sc}(T)}{\sigma_{sc}(0)} = 1 + 2 \left\langle\int_{\Delta_k}^{\infty}\frac{\partial f}{\partial E}\frac{E{\rm d}E}{\sqrt{E^2-\Delta_k^2}}\right\rangle_{\rm FS},
\label{RhoS}
\end{equation}

\noindent where $f = [1+\rm{exp}(-E/k_B T)]^{-1}$ is the Fermi function, $\varphi$ is the azimuthal angle along the Fermi surface and the brackets correspond to an average over the Fermi surface. The gap is given by $\Delta(T, \varphi)$=$\Delta_0 \delta(T/T_C)g(\varphi)$, where the temperature dependence  is given by $\delta(T/T_c)$ =tanh$[(\pi T_c/\Delta_0)\sqrt{(2/3)(\Delta C/\gamma T_c)(T_c/T-1)}]$~\cite{UBe13}, with $\Delta C/\gamma T_c = 2.2$, \cite{T. Kong} and g($\varphi$)=1 for $s$-wave model and g($\varphi$)=$|\cos(2\varphi)|$ for a $d$-wave model with line nodes.

\begin{figure}[t]
\vskip -0 cm
\centering
\includegraphics[width = 7 cm]{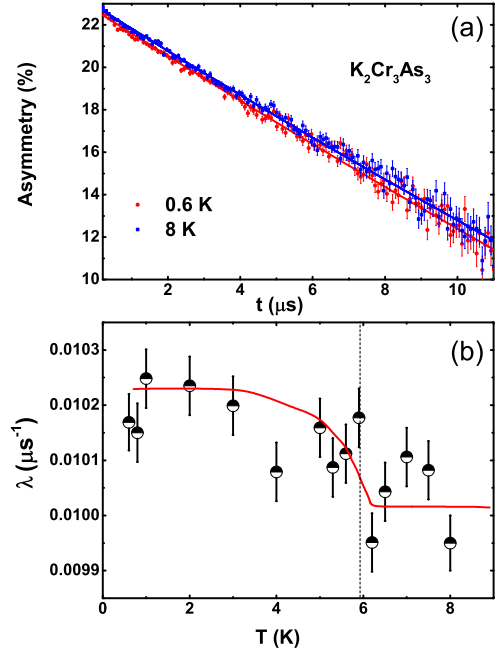}
\caption {(Color online) (a) Zero-field $\mu$SR time spectra for K$_2$Cr$_3$As$_3$ collected at 0.6 K (red circle) and 8.0 K (blue square) are shown together with lines that are least squares fits to the data using Eq. (3). These spectra collected below and above $T_{\bf c}$ are representative of the data collected over a range of  $T$. (b) The temperature dependence of the electronic relaxation rate measured in zero magnetic field of K$_2$Cr$_3$As$_3$ with $T_{\bf c}$ = 5.8 K is shown by the vertical dotted line. The solid red line is guide to the eye.}
\end{figure}

\par
Fig. 3 (a) shows the $T$ dependence of the $\sigma_{sc}$, measured in an applied field of 400 G through two different modes: zero-field-cooled (ZFC) and  field cooled (FC). The $\sigma_{sc}$ can be directly related to the superfluid density. The temperature dependence of  $\sigma_{sc}$ shows the establishment of a flux-line lattice and indeed indicates a decrease of the magnetic penetration depth with decreasing temperature. Comparing the ZFC and FC data reveals a substantial difference. In the ZFC mode, $\sigma_{sc}$ increases with decreasing temperature faster than that of FC, thus points to differences in the numbers of the pinning sites and trapping energies, which are altered by magnetic fields and sample history. From the observed temperature dependence of $\sigma_{sc}$, the nature of the superconducting gap can be determined.  The temperature dependence of $\sigma_{sc}$  shows evidence of saturation type behaviour at the lowest temperature, which is a typical behaviour observed for a fully gapped superconductor. For example $\sigma_{sc}$ of Li$_2$Pd$_3$B exhibits very similar behavior and the results have been interpreted as due to a single $s$-wave gap~\cite{LiPdB}.  The $\sigma_{sc}(T)$ data of  K$_2$Cr$_3$As$_3$ can be well modeled by a single isotropic gap of 0.80(5)~meV using Eq.(2) (see Fig.3b). This gives a value of $\Delta_0$/$k_B$$T_{\bf c}$ = 1.6(1), which is slightly lower than the 1.764 expected for BCS superconductors, which may be due to an anisotropic gap. Considering the tunnel diode study reveals a possibility of line nodes~\cite{G. M. Pang}, we have also fitted  $\sigma_{sc}$ data using a $d$-wave model with line nodes. The fit to the nodal model (solid line Fig.3b) shows similar agreement to that of $s$-wave (short-dash line Fig.3b), but gives a higher value of $\Delta_0$/$k_B$$T_{\bf c}$ = 3.2$\pm$0.2. We also tried $d$-wave fits with various fixed value of $\Delta_0$/$k_B$$T_{\bf c}$ and allowing only $\sigma_{sc}(0)$  to vary. Our results show that $~\chi^2$ increase dramatically for lower values of $\Delta_0$/$k_B$$T_{\bf c}$ (see the inset in Fig.3b, and dotted line). We therefore believe that the gap function of K$_2$Cr$_3$As$_3$ can   be equally accounted for either with an  $s$-wave or $d$-wave model. It is to be noted that for Li$_2$Pt$_3$B the  $\mu$SR study reveals a single-gap isotropic $s$-wave behavior~\cite{LiPdB}, but the tunnel diode as well as NMR studies reveal a spin-triplet pairing with line nodes in the gap function ~\cite{LiPtBTD, LiPtBNMR}.

The observed increase in $\sigma_{sc}$ of K$_2$Cr$_3$As$_3$  gives rise to the large value of the muon spin depolarization rate below superconducting transition temperature and is related to the magnetic penetration depth. For a triangular lattice,~\cite{jes,amato,chia} $\frac{\sigma_{sc}(T)^2}{\gamma_\mu^2}= \frac{0.00371\phi_0^2}{\lambda^4(T)}$, where $\gamma_\mu/2\pi$ = 135.5 MHz/T is the muon gyromagnetic ratio and $\phi_0$ = 2.07$\times$10$^{-15}$ T m$^2$ is the flux quantum. As with other phenomenological parameters characterizing a superconducting state, the penetration depth can also be related to microscopic quantities. Using London theory~\cite{jes}, $\lambda_L^2= m^{*}c^2/4\pi n_s e^2$, where m$^*$ = (1+$\lambda_{e-ph}$)m$_e$ is the effective mass and $n_s$ is the density of superconducting carriers. Within this simple picture, $\lambda_L$ is independent of magnetic field. $\lambda_{e-ph}$ is the electron-phonon coupling constant that can be estimated from $\Theta_D$ and $T_c$ using McMillan's relation~\cite{mcm} $\lambda_{e-ph}=\frac{1.04+\mu^*ln(\Theta_D/1.45T_{\bf c})}{(1-0.62\mu^*)ln(\Theta_D/1.45T_{\bf c})+1.04}$, where $\mu^*$ is the repulsive screened Coulomb parameter usually assigned as $\mu^*$ = 0.13. For K$_2$Cr$_3$As$_3$ we have $T_{\bf c}$ = 5.8 K and $\Theta_D$ = 216 K, which together with $\mu^*$ = 0.13, we have estimated $\lambda_{e-ph}$ = 0.75. K$_2$Cr$_3$As$_3$ is a type II superconductor, assuming that roughly all the normal state carriers ($n_e$) contribute to the superconductivity (i.e., $n_s\approx n_e$), hence  we have estimated the magnetic penetration depth $\lambda$, superconducting carrier density $n_s$, and effective-mass enhancement $m^*$  to be $\lambda_L(0)$ = 454(4) nm (from the $s$-wave fit), $n_s$ = 2.4$\times$10$^{27}$ carriers/m$^3$, and $m^*$ = 1.75 $m_e$, respectively. More details on these calculations can be found in Ref.~\cite{adsd,vkasd,dtasd}.

\par
The time evolution of the ZF$-\mu$SR is shown in Fig. 4 (a) for $T$ = 600 mK and 8 K. In these relaxation experiments, any muons stopping on the titanium sample holder give a time independent background. No signature of precession is visible, ruling out the presence of a sufficiently large internal magnetic field as seen in magnetically ordered compounds. One possibility is that the muon$-$spin relaxation is due to static, randomly oriented local fields associated with the nuclear moments at the muon site. The ZF$-\mu$SR data are well described by,

\begin{equation}
G_{z2}(t) =A_1 e^{-\lambda t}+A_{bg}
\end{equation}

\noindent where $\lambda$ is the electronic relaxation rate, $A_1$ is the initial asymmetry, $A_{bg}$ is the background.  The parameters $A_1$, and $A_{bg}$ are found to be temperature independent. It is remarkable that $\lambda$ shows a moderate increase [Fig. 4 (b)] with an onset temperature of $\ge$ 6.0$\pm$0.1 K, indicating the appearance of a spontaneous internal field or slow down of spin fluctuations correlated with the superconductivity.  Based on this evidence, we propose that TRS might be broken in the SC state of K$_2$Cr$_3$As$_3$. Such a change in $\lambda$ has only been observed in superconducting Sr$_2$RuO$_4$~\cite{gm}, LaNiC$_2$~\cite{ad1} and Lu$_5$Rh$_6$Sn$_{18}$~\cite{ab1,ab2}.  This increase in $\lambda$  can be explained in terms of a signature of a coherent internal field with a very low frequency as discussed by Luke {\it et. al.}~\cite{gm} for Sr$_2$RuO$_4$. This suggests that the field distribution is Lorentzian in nature similar to Sr$_2$RuO$_4$ with an average of the second moment of the field distribution of  0.003 G in K$_2$Cr$_3$As$_3$. This value is very small compared to 0.5 G observed in Sr$_2$RuO$_4$~\cite{gm}. 

In summary, we have carried out zero-field (ZF) and transverse field (TF) muon spin rotation ($\mu$SR) experiments in the superconducting state of  K$_2$Cr$_3$As$_3$, which has a quasi-one-dimensional crystal structure.  Our ZF $\mu$SR data reveal the presence of a very weak internal field (0.003 G) or slow down of spin fluctuations in the superconducting state. From the TF $\mu$SR we have determined the muon depolarization rate in ZFC and FC modes associated with the vortex-lattice.  Furthermore,  the temperature dependence of $\sigma_{sc}$  can be fitted equally well to either a single-gap isotropic $s-$wave or a $d$-wave model with line nodes. Considering the possible multi-band nature of superconductivity in K$_2$Cr$_3$As$_3$ one would expect more complex behavior of the gap function and hence the conclusions obtained from our TF $\mu$SR study are in line with this.  The line nodes in the superconducting gap need further investigations of $\mu$SR on a good quality single crystals of K$_2$Cr$_3$As$_3$.

DTA and ADH would like to thank CMPC-STFC, grant number CMPC-09108, for financial support. A.B would like to acknowledge FRC of UJ, NRF of South Africa and ISIS-STFC for funding support. Work at Fudan University is supported by the Shanghai Pujiang Scholar Program (Grant No.13PJ1401100). AMS thanks the SA-NRF (Grant 93549) and UJ Research Committee for  financial support.

\end{document}